\documentclass{article}
\usepackage{unicode-math}
\usepackage{graphicx}
\usepackage[margin=1in]{geometry}
\usepackage{amsmath, amsthm, bbm, braket, complexity}
\usepackage{xcolor}
\usepackage{xr}
\usepackage{hyperref}
\usepackage{cleveref}
\usepackage{parskip}
\usepackage{booktabs}
\usepackage{multirow}
\usepackage{caption}
\usepackage{subcaption}
\usepackage{authblk}
\usepackage{svg}
\usepackage{framed}
\usepackage[maxnames=3,minnames=3,sorting=none,bibstyle=numeric-comp]{biblatex}
\RequirePackage[title,titletoc]{appendix}

\usepackage{xfp}
\newcommand\SupplementaryMaterials{%
  \xdef\presupfigures{\arabic{figure}}
  \xdef\presupsections{\arabic{section}}
  \renewcommand\thefigure{S\fpeval{\arabic{figure}-\presupfigures}}
  \renewcommand\thesection{S\fpeval{\arabic{section}-\presupsections}}
}

\newcommand{\ie}{\emph{i.e.\ }}

\newtheorem{problem}{Problem}
\crefname{problem}{problem}{problems}

\hypersetup{ 
  pdftitle={Pangenome-guided sequence assembly via binary optimisation},
  pdfauthor={Josh Cudby, James Bonfield, Chenxi Zhou, Richard Durbin, Sergii Strelchuk}, 
  pdfsubject={Methods for solving the problem of assembling sequences from a pangenome and a set of short reads}, 
  pdfkeywords={Pangenomics, Genomics, Bioinformatics, Optimisation, Quantum}, 
  pdfdisplaydoctitle=true 
}

\title{Pangenome-guided sequence assembly via binary optimisation}

\author[1,4]{Josh Cudby}
\author[2]{James Bonfield}
\author[3]{Chenxi Zhou}
\author[3]{Richard Durbin}
\author[4]{Sergii Strelchuk}

\date{\today}

\affil[1]{\small Department of Applied Mathematics and Theoretical Physics, University of Cambridge, Wilberforce Rd, Cambridge CB3 0WA, United Kingdom (jjcc2@cam.ac.uk)}
\affil[2]{Wellcome Sanger Institute, Hinxton, Cambridge CB10 1RQ, United Kingdom}
\affil[3]{ Department of Genetics, University of Cambridge, Downing Street, Cambridge CB2 3EH,
United Kingdom}
\affil[4]{Department of Computer Science, University of Oxford, Parks Rd, Oxford OX1 3QG, United Kingdom}

\begin{document}

\maketitle

\begin{abstract}
\emph{De novo} genome assembly is challenging in highly repetitive regions; however, reference-guided assemblers often suffer from bias.
We propose a framework for pangenome-guided sequence assembly, which can resolve short-read data in complex regions without bias towards a single reference genome.
Our primary contribution is to frame the assembly as a graph traversal optimisation problem, which can be implemented classically or on a quantum computer.
The workflow involves first annotating pangenome graphs with estimated copy numbers for each node, then finding a path on the graph that best explains those copy numbers.
On simulated data, our approach significantly reduces the number of contigs compared to \emph{de novo} assemblers. 
While they introduce a small increase in inaccuracies, such as false joins, our optimisation-based methods are competitive with current exhaustive search techniques. 
They are also designed to scale more efficiently as the problem size grows and will run effectively on future quantum computers; a small experiment on a real quantum device showcases this behaviour.
Moreover, they are more resilient to noise in copy number estimation inherent in short-read-based assembly.
We also develop novel tools for creating realistic synthetic pangenomes, aligning reads to pangenomes and for evaluating assembly quality.
\end{abstract}


\textbf{Authors:} all authors are part of a research collaboration participating in the Wellcome Leap Q4Bio programme, of which Sergii Strelchuk is the main PI and Richard Durbin a co-PI.

\section{Introduction}
A single linear reference genome alone cannot capture the genome variation that occurs within species, whether the variation is local, such as Single Nucleotide Variants (SNVs), or larger structural changes, like Copy Number Variations (CNVs)~\cite{ballouz_is_2019}.
Bioinformatic pipelines often include a step where a new sample is compared to this fixed reference genome; for example, in variant detection.
When a single reference is used, \emph{reference bias} is inevitable: a tendency to report fragments of sequence that appear in the reference, and to miss novel or highly divergent regions~\cite{ballouz_is_2019,huang_short_2013,gage_multiple_2019}.
The solution is to use references that capture genomic diversity; one such object is the pangenome.

Pangenomics is a relatively young field, and there is no unified description of the various forms and construction methods of pangenomes.
Indeed, the term ``pangenome'' is also applied to representations that separate the core and accessory genes present in bacterial species~\cite{tettelin_genome_2005}, a completely different setting than we are interested in.
According to the conventions set out by Matthews et al.~\cite{matthews_gentle_2024}, we focus on \emph{sequence-oriented pangenome graphs}, which we henceforth refer to simply as pangenomes or pangenome graphs.
These objects describe the location and nature of genomic variation within a collection of individuals, most commonly from the same species.
As the name suggests, they are described by mathematical graphs consisting of a set of \emph{nodes}, representing fragments of sequence, connected by \emph{edges}, representing fragments that are adjacent within a genome.
These graphs can be created by popular tools such as \emph{vg}~\cite{hickey_genotyping_2020}, \emph{pggb}~\cite{liao_draft_2022} and \emph{minigraph}~\cite{li_design_2020}.
The Human Pangenome Reference Consortium (HPRC) recently published an initial \emph{human} pangenome~\cite{liao_draft_2022} containing data from 47 individuals, with a follow-up data release of more than 200 genomes available online.

Due to dramatic reductions in the cost of next-generation sequencing, we now have whole-genome shotgun short-read data for millions of human genomes.
Using a pangenome to reconstruct the full sequences of these samples would be preferable to using standard references such as GRCh38~\cite{schneider_evaluation_2017} or performing incomplete \emph{de novo} assembly.
A toolkit dedicated to providing this functionality is not currently available, though some software kits use pangenomes in other ways.
\textit{DRAGEN}~\cite{behera_comprehensive_2025} performs variant detection by mapping reads against a pangenome before mapping back to a comparison against a single reference genome.
\textit{Pasa}~\cite{do_pasa_2024} uses pangenomes as a source of global information to resolve assemblies with a large number of contigs into a single sequence.

\begin{figure}
    \centering
    \includegraphics[width=0.95\textwidth,alt={A flow chart showing the steps involved in pangenome-guided sequence assembly.}]{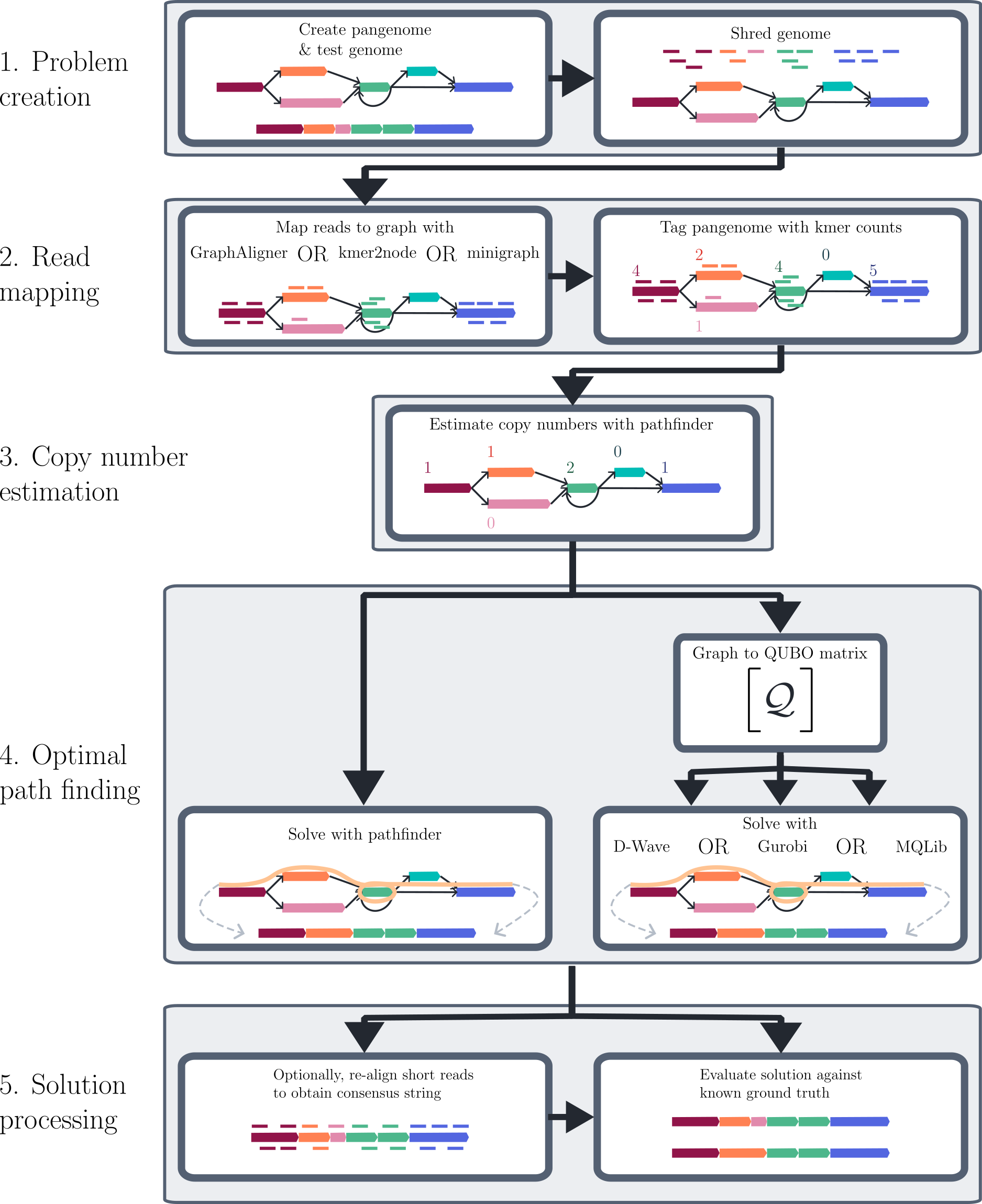}
    \caption{
    A sketch of the pangenome-guided sequence assembly procedure. 
    (1) \textit{Problem creation} consists of two steps. A pangenome and a new individual genome are synthesised, and the genome is shredded to simulate shotgun sequencing. 
    (2) \textit{Read mapping} involves aligning the short reads to the pangenome using 1 of 3 software tools and tagging the nodes with the observed kmer counts. 
    (3) \textit{Copy number estimation} is performed using \textit{pathfinder}.
    (4) \textit{Optimal path finding} depends on the choice of solver. If using \textit{pathfinder}, input the annotated graph directly. Otherwise, construct the QUBO matrix from the graph, and input that into the chosen QUBO solver.
    (5) \textit{Solution processing} starts with an optional re-alignment step to improve solution quality, before evaluating solution quality with a variety of metrics.
    }
    \label{fig:pangenome_assembly}
\end{figure}

Here, we propose a methodology for \textit{pangenome-guided sequence assembly} that maps reads onto a pangenome and reformulates the assembly as a graph-traversal optimisation.
Most genomic regions are easily resolved; we focus on the highly complex, repetitive regions of the pangenome graph.
The pipeline, summarised in~\cref{fig:pangenome_assembly}, consists of five stages:
\begin{description}
    \item[1. Problem creation:] Generate a synthetic pangenome and a new individual genome. Simulate short-read sequencing (typically 30× coverage).
    \item[2. Read mapping:] Align short reads to the pangenome using \textit{GraphAligner}, \textit{kmer2node} or \textit{minigraph}, and annotate nodes with kmer counts.
    \item[3. Copy number estimation:] Estimate node copy numbers from the annotated graph.
    \item[4. Path finding:] Identify a path through the pangenome graph that best fits the observed copy numbers using either the \textit{pathfinder} tool or a novel binary optimisation approach.
    \item[5. Solution processing:] Extract and optionally re-align the resulting sequence to improve quality before evaluating against the ground truth.
\end{description}

The key computational task in the pipeline is the path finding in step 4. This problem is similar to those tackled by certain \textit{de novo} genome assemblers like \textit{Flye}~\cite{kolmogorov_assembly_2019} and \textit{Oatk}~\cite{zhou_oatk_2024}, which aim to resolve repeat-induced non-linear structures in assembly graphs. 
Their approach typically involves estimating the copy numbers of repeat sequences from sequence coverage and then identifying a graph traversal that satisfies these copy number constraints. 
They are designed for datasets derived from a single individual, in which the resulting graphs are typically complete and relatively noise-free, allowing the estimation of sequence copy numbers reliably.

In the context of pangenome graphs, however, additional complexities arise. 
First, due to genomic divergence, a pangenome graph does not fully represent the target genome, necessitating imputation of missing or divergent regions.
Second, read mapping to a pangenome graph is inherently noisier due to its repetitive and multi-genome nature, resulting in greater uncertainty in sequence coverage and, consequently, in copy number estimates. 
To accurately resolve the graph paths through repetitive regions, pangenome-based methods must be more robust to such noise.

To that end, we formalise the path finding problem as an optimisation problem. 
The optimisation may be viewed as a generalisation of the Hamiltonian path problem and is $\NP$-hard in general.
We cast it as a Quadratic Unconstrained Binary Optimisation (QUBO), a framework commonly used in the optimisation literature due to its ability to strike a balance between expressivity and tractability. 
Moreover, QUBO is in some sense ``native'' for \emph{quantum} optimisation, particularly the Quantum Adiabatic Optimisation Algorithm (QAOA).
Quantum computers are currently limited in size and error-prone, characteristic of the so-called Noisy Intermediate-Scale Quantum (NISQ) era.
In the NISQ era, applications of quantum computers must be carefully chosen to mitigate these limitations; combinatorial optimisation is a promising direction that has already seen some  success~\cite{kim_quantum_2025,mohanty_analysis_2023,perez-ramirez_variational_2024,sachdeva_quantum_2024}.
A primer on quantum computing and quantum optimisation is given in~\cref{sec:quantum}.
Below, we compare both classical and quantum approaches to the optimisation and discuss the potential utility of quantum approaches.

We also make several contributions that may be of interest to the broader (pan)genomics community.
We develop a tool for creating synthetic pangenomes that have a realistic biological structure.
The size and complexity of these pangenomes can be controlled by adjusting the rate of simulated evolutionary mutations, such as CNVs, large repeats and translocations.
We also create a novel tool \emph{kmer2node} for aligning reads onto pangenome graphs.
Finally, we develop a post-processing pipeline to evaluate assembly quality and improve it with an optional step.
\section{Results}
Our main comparison point for existing state-of-the-art is \emph{pathfinder}~\cite{zhou_oatk_2024}. 
\emph{Pathfinder} is a submodule from the \emph{Oatk} package, a tool for resolving plant organelles using high-accuracy long reads.
While this application is far from our current setting, the methods are somewhat related.
In particular, \emph{pathfinder} also assigns weights to a graph and then searches for a walk that explains the data.
However, it proceeds via exhaustive search rather than optimisation; moreover, \textit{pathfinder} was designed to resolve assembly graphs whose annotated coverage values are essentially noiseless and have no unused nodes.
Conversely, in our setting, the reads originate from an individual \textit{not} present in the pangenome, and we therefore expect imperfect coverage values and many nodes to have zero coverage.
We use a cost function to be robust to this noise.

As an example, consider a simple line graph with nodes $A \rightarrow B \rightarrow C \rightarrow D \rightarrow E$ and estimated copy numbers $(1,\, 1,\, 0,\, 1,\, 1)$.
The prescriptive nature of \textit{pathfinder} yields two contigs: $A \rightarrow B$ and $C \rightarrow D$.
Conversely, the cost function used in our QUBO formulation allows a single unified path of $A \rightarrow \ldots \rightarrow E$ to be found. 
Nevertheless, to the best of our knowledge, \textit{pathfinder} is the software package that gives the fairest comparison with our results.

A comparison of pangenome-guided assembly of synthetic haploid data via kmer mapping using \emph{pathfinder} against \emph{de novo} sequence assembly can be seen in~\cref{tab:denovoassembly}.
The primary difference lies in the number of contigs returned and demonstrates the great benefit of using a pangenome-guided approach to short-read assembly.
However, this comes with a small increase in false joins (\textit{i.e.}, breaks in alignment with the truth set).
Optimising the path sequence by realigning the short-read data against it and producing a new consensus improves accuracy across a range of metrics.
The percentage identity remains slightly lower than that of the \emph{de novo} assemblers, indicating that a hybrid solution may be the optimal strategy.

\begin{table}
{\begin{tabular}{@{}lrrrrrrrr@{}}
\toprule
Annotator/Assembler & \%Covered & \%Used & Contigs & Breaks & Indel & No.Diff & N50 & \%Identity \\ \midrule
Kmer2node        &  88.3  &  93.2  &  2.8   &  1.9  &  1.2  &  0.2  &  6697 & 96.9 \\
Minigraph        &  85.1  &  91.2  &  4.7   &  2.3  &  1.0  &  \textbf{0.0}  & 5044 &  97.2 \\
GraphAligner     &  92.6  &  92.0  &  1.7   &  2.0  &  1.7  &  0.2  & 9484 &  96.9 \\
Giraffe          &  91.7  &  91.7  &  1.9   &  2.0  &  1.4  &  0.1  & 8748 &  97.0 \\ \midrule
Kmer2node-opt    &  88.4  &  \textbf{94.0}  &  2.6  &  1.5  &  0.6  & 0.3 & 6679 &  98.7 \\
Minigraph-opt    &  85.4  &  92.9  &  4.3   &  1.8  &  0.3  &  0.1  &  5002 & 99.2 \\
GraphAligner-opt &  \textbf{93.0}  &  92.6  &  \textbf{1.6}  &  1.6  &  0.8  &  0.3 & \textbf{9436} &  98.6 \\
Giraffe-opt      &  91.9  &  92.6  &  1.8  &  1.7  &  0.7  &  0.3 & 8701 &  98.7 \\ \midrule
Syncasm          &  91.6  &  86.2  &  31.3  &  \textbf{0.0}  &  \textbf{0.1}  &  \textbf{0.0} & 878 &  \textbf{100.0}\\
Miniasm          &  81.5  &  91.3  &  14.8  &  0.1  &  \textbf{0.1}  &  \textbf{0.0}  & 1022 &  99.6 \\ \bottomrule
\end{tabular}}
\caption{Performance of pangenome guided mapping of short-read data using \emph{minigraph}, \emph{GraphAligner}, \emph{vg giraffe} and \emph{kmer2node} weight assignments with the \emph{pathfinder} solver, compared against pure \emph{de novo} sequence assembly from \emph{syncasm} and \emph{miniasm}. 
Weight assignment methods are discussed in detail in~\cref{sec:node_weight}. 
Averaged stats from 51 seeds with 5 alignments each.
\emph{Pathfinder} results are presented twice, evaluating the original concatenated path and an optimised path formed by realigning all the short-read data back to the original path to produce a refined consensus sequence. 
Columns represent the proportion of the true genome covered by the assembly result (\%Covered), and the percentage of this result covered by the true genome (\%Used), the number of contigs, the number of breaks in alignment of assembly to true sequence (breaks, \ie false joins), the number of large indels and regions of significant runs of differences, the N50 contig size, and total percent identity between assembly and true genome. 
See~\cref{sec:classical_post} for further details.
}
\label{tab:denovoassembly}
\end{table}

To evaluate the optimisation formulation, we tested a range of solvers with varying levels of quantum integration. 
These methods span a spectrum from traditional classical solvers, to hybrid approaches like quantum annealing QA, which is specialised for optimisation problems, to gate-based quantum computers.

Classical solvers and quantum annealers can accept large problems with up to several thousand variables.
However, they do not necessarily find optimal or even good solutions within a reasonable time frame at this scale.
Conversely, the quantum circuit approach is very limited in scope.
Quantum optimisation experiments have been run on hardware with up to 127 qubits~\cite{sachdeva_quantum_2024}.
In this work, we limit ourselves to simulations of such experiments.
Due to the exponentially scaling cost of simulating quantum systems as the number of qubits increases, we only consider problems with up to 35 qubits. 
The results of these simulations serve as a proof-of-principle for a quantum approach as hardware size and quality improve in the coming years.

\subsection{Classical optimisers}

We first benchmark the performance of our formulation using two classical optimisation suites.
\emph{Gurobi}~\cite{gurobi_optimization_llc_gurobi_nodate} is an industry-standard package that employs branch-and-bound methods to find provably optimal solutions to a range of optimisation problems, including mixed-integer quadratic programming tasks which admit QUBO as a special case.
\emph{MQLib}~\cite{dunning_what_2018} is a package that provides fast implementations of several heuristics for QUBO.
We use the multistart tabu search strategy introduced by Palubeckis~\cite{palubeckis_multistart_2004} since it performs most consistently on our instances.

We test the performance of each solver -- \emph{pathfinder}, \emph{Gurobi} and \emph{MQLib} -- on graphs obtained with each annotation strategy -- \emph{GraphAligner}, \emph{minigraph} and the in-house \emph{kmer2node} tool.
We use the Oriented Tangle Resolution problem, defined in~\cref{sec:assembly2opt}, as it captures the biological nature of the task while having reduced complexity compared to the Diploid version, making it easier to benchmark across all the solution strategies. 

\begin{figure}
    \centering
    \includegraphics[alt={A set of radar charts and violin plots comparing the relative performance of combinations of annotation strategies and classical solvers.}]{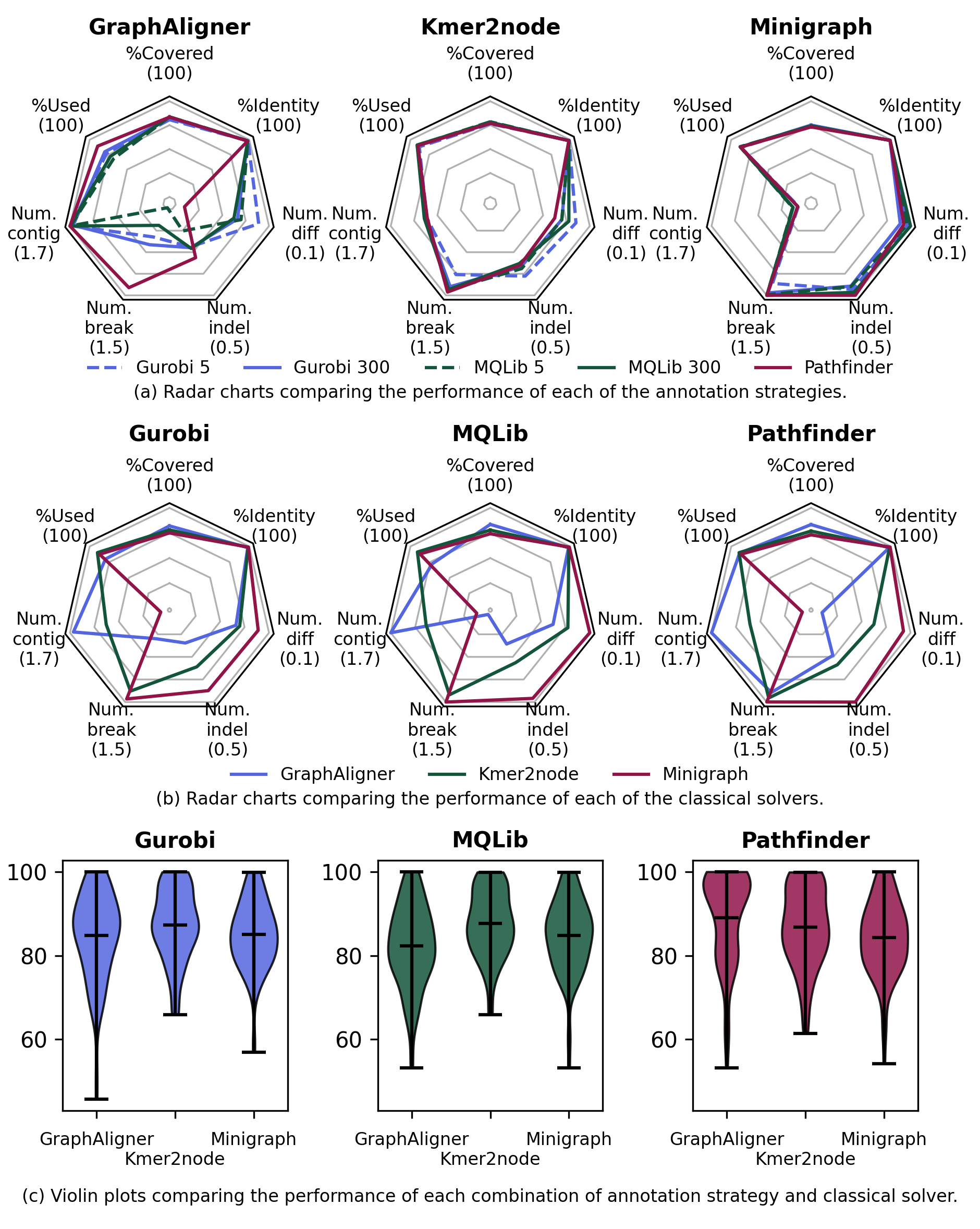}
        \caption{(a), (b) Radar charts comparing the performance of each combination of annotation strategy: \emph{GraphAligner}, \emph{kmer2node} and \emph{minigraph}; and classical solver: \emph{Gurobi}, \emph{MQLib} and \emph{pathfinder}. The seven axes plotted are the evaluation criteria discussed in~\cref{sec:classical_post}. For each, the number in brackets corresponds to the outermost point on that axis. For the lower half of each plot, further out along the axis corresponds to fewer contigs, breaks and so on.
        (c) Violin plots showing the per-instance performance of \emph{Gurobi}, \emph{MQLib} and \emph{pathfinder}.}
        \label{fig:compare_classical}
\end{figure}

We test each combination on a set of 20 pangenomes with 5 sequences to be aligned against each pangenome.
The pangenomes have an average of 77.8 nodes.
\emph{Gurobi} and \emph{MQLib} are given time limits of 5 and 300 seconds to test whether higher quality solutions are obtained at long run times.
These classical optimisers are allowed 3 runs per sequence and per time limit to account for variations in solution quality due to their heuristic nature.
The averaged results are given in~\cref{tab:classical_solvers}.
Radar charts showing the average performance across the metrics discussed in~\cref{sec:classical_post} are given in~\cref{fig:compare_classical}.
\Cref{fig:compare_classical}a shows that, for any choice of annotation strategy, each of the solvers is broadly competitive with the others. 
When either \emph{kmer2node} or \emph{minigraph} is used, the average case results across each of the solvers are very similar.
The starkest difference comes when \emph{GraphAligner} is used.
In that case, \emph{pathfinder} reports more large differences than either \emph{MQLib} or \emph{Gurobi}, but fewer breaks and slightly fewer large indels on average.
The large number of breaks in the QUBO solver results may be due to not taking into account edge weights, which can help to resolve inversions: see~\cref{sec:discussion} for a brief discussion.

\begin{table}
\begin{tabular}{@{}llrrrr@{}}
\toprule
Annotator & Solver & \%Cov. & \%Used & Contigs & N50\\ \midrule
\multirow{3}{*}{GraphAligner}
& Gurobi & 85.69 (12.46) & 83.85 (11.75) & 1.92 (1.52) & \textbf{9755} (2699)     \\
& MQLib & \textbf{86.90} (12.97) & 77.75 (13.60) & \textbf{1.73} (1.13) & 11337 (3846) \\
 & Pathfinder & 86.74 (15.82) & 91.42 (8.56) & \textbf{1.73} (1.13) & 8725 (2194) \\ \midrule
\multirow{3}{*}{Kmer2node} 
& Gurobi & 82.50 (12.52) & 92.07 (7.68) & 3.73 (2.50) & 5370 (2529)\\
& MQLib & 82.36 (12.19) & \textbf{92.88} (6.44) & 3.68 (2.27) & 5201 (2523) \\
 & Pathfinder & 81.55 (13.42) & 92.01 (7.76) & 3.85 (2.71) & 5208 (2586)  \\ \midrule
\multirow{3}{*}{Minigraph} 
& Gurobi & 80.09 (11.74) & 89.92 (7.37) & 6.74 (3.69) & 3512 (2708) \\
& MQLib & 79.40 (13.45) & 90.31 (7.59) & 6.47 (3.58) & 3468 (2696) \\
 & Pathfinder & 78.51 (13.84) & 90.06 (7.48) & 6.74 (3.84) & 3409 (2684) \\ \bottomrule
\end{tabular}
\\[12pt]
\begin{tabular}{@{}llrrrr@{}}
\toprule
Annotator & Solver & Breaks & Indels & Diffs & \%Identity \\ \midrule
\multirow{3}{*}{GraphAligner}
& Gurobi      & 5.31 (3.76) & 1.09 (1.09) & 0.18 (0.48) & 98.45 (1.38)   \\
& MQLib       & 6.76 (4.90) & 1.08 (1.14) & 0.19 (0.44) & 98.42 (1.35) \\
 & Pathfinder & 2.09 (2.08) & 0.97 (1.18) & 0.33 (0.62) & 98.21 (1.90)    \\ \midrule
\multirow{3}{*}{Kmer2node} 
& Gurobi      & 2.17 (2.09) & 0.86 (1.04) & 0.17 (0.40) & 98.82 (1.14)     \\
& MQLib       & 1.95 (1.72) & 0.90 (1.14) & 0.15 (0.36) & 98.75 (1.32) \\
 & Pathfinder & 1.76 (1.53) & 0.88 (1.05) & 0.19 (0.42) & 98.70 (1.34) \\ \midrule
\multirow{3}{*}{Minigraph} 
& Gurobi      & 1.70 (1.58) & 0.63 (0.95) & 0.12 (0.35) & 99.09 (0.93)   \\
& MQLib       & \textbf{1.52} (1.40) & 0.56 (0.82) & \textbf{0.09} (0.29) & 99.14 (0.75)  \\
 & Pathfinder & \textbf{1.52} (1.34) & \textbf{0.52} (0.79) & 0.11 (0.34) & \textbf{99.17} (0.91) \\ \bottomrule
\end{tabular}
\caption{Performance of pangenome guided mapping of short-read data using \emph{minigraph}, \emph{GraphAligner} and \emph{kmer2node} weight assignments with the \emph{MQLib}, \emph{Gurobi} and \emph{pathfinder} solvers. Averaged stats from 20 seeds with 5 alignments each, with standard deviations given in parentheses. \emph{MQLib} and \emph{Gurobi} were given 300 seconds per run, and 3 runs per alignment. }
\label{tab:classical_solvers}
\end{table}

As well as average-case performance, it is illuminating to see how the combinations perform on specific instances, particularly hard ones.
Violin plots showing the average of the ``Covered'' and ``Used'' statistics across all instances are given in~\Cref{fig:compare_classical}c.
When \emph{GraphAligner} is used to annotate graphs, it is clear that \emph{pathfinder} outputs high-quality solutions on a majority of instances, although a very low worst-case outcome drags down its average performance.
For the other annotators, the QUBO solvers are much more competitive, with similarly shaped plots across both the \emph{minigraph} and \emph{kmer2node} settings.

\subsection{Quantum annealing}

To test the potential of QA, we perform a range of experiments using the \emph{Leap} hybrid solver provided by the quantum hardware company \emph{D-Wave}~\cite{daigle_d-wave_2025}.
This solver is provided on the cloud, and typically one or more classical algorithms run on the problem while hard parts are outsourced to the QPU.
A configurable time limit is provided, and the best solution found during that time is returned, along with some metadata. 
The main benefit of QA is the size of the hardware available. 
The Advantage2 system has 4,400 qubits arranged in a lattice structure, around 30 times larger than superconducting offerings from IBM and around 90 times larger than Quantinuum's trapped ion computers.
We use the QA systems to wayfind the potential for quantum advantage on more standard quantum hardware in the future.

\begin{figure}
    \centering
    \includegraphics[alt={A set of radar charts and violin plots comparing the relative performance of combinations of annotation strategies and either the D-Wave or pathfinder solvers.}]{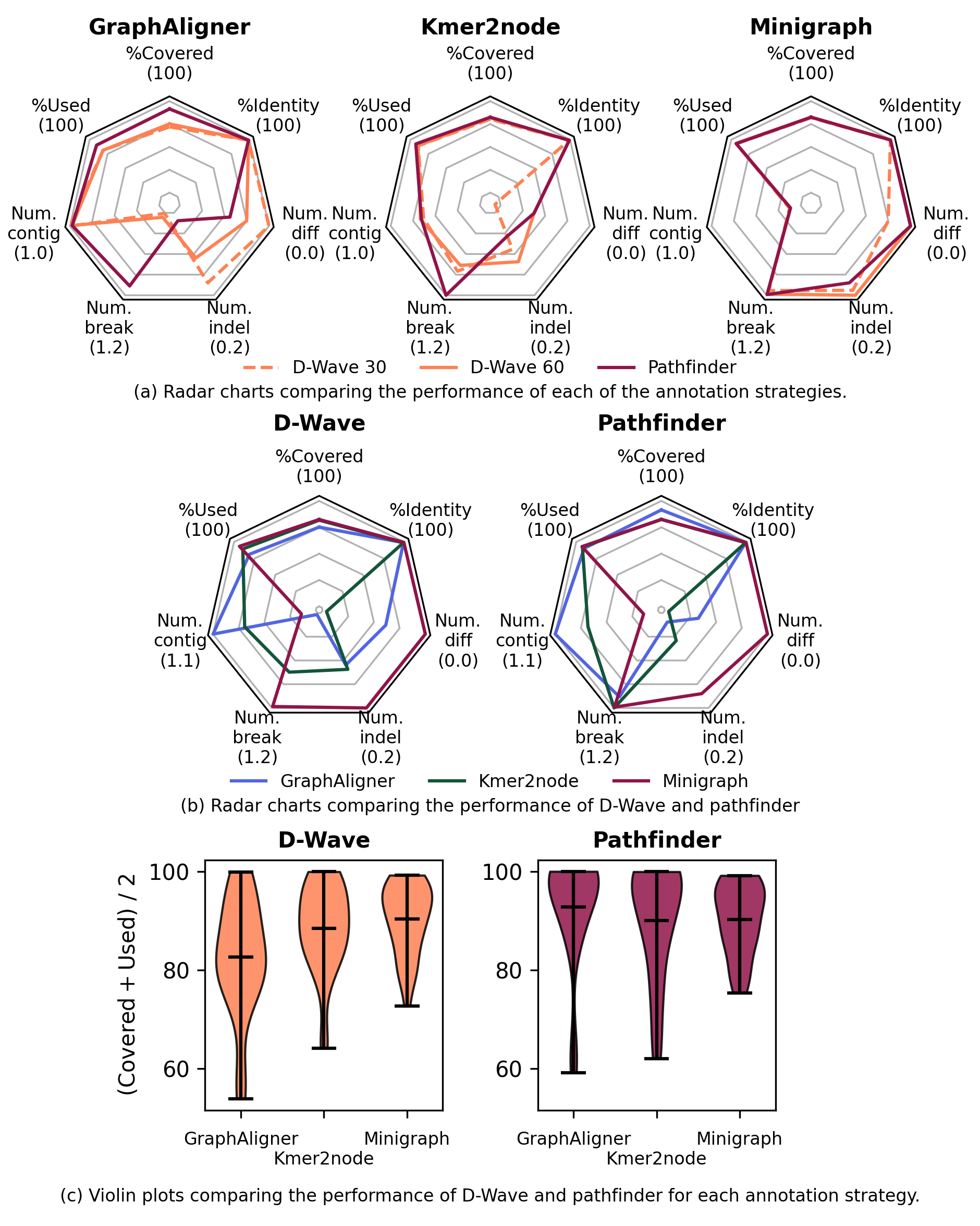}
    \caption{(a), (b) Radar charts comparing the performance of each combination of annotation strategy: \emph{GraphAligner}, \emph{kmer2node} and \emph{minigraph}; with quantum annealing solver \emph{D-Wave} or classical solver \emph{pathfinder}. The seven axes plotted are the evaluation criteria discussed in~\cref{sec:classical_post}. For each, the number in brackets corresponds to the outermost point on that axis. For the lower half of each plot, further out along the axis corresponds to fewer contigs, breaks and so on. (c) Violin plots showing the per-instance performance of \emph{D-Wave} and \emph{pathfinder}.} 
    \label{fig:compare_dwave}
\end{figure}

We test \emph{D-Wave} on 10 pangenomes, with 3 sequences to be aligned to each. 
Due to limited access to \emph{D-Wave} systems, we tested on smaller instances with an average of 44.4 nodes per graph.
The annealer was allowed 2 runs per sequence, with a time limit of either 30 or 60 seconds.
Results are given in~\cref{tab:dwave_solvers} and visualised in~\cref{fig:compare_dwave}.
For the instances tested, there is a clear benefit to using \emph{minigraph} as an annotator; in that case, \emph{D-Wave} performs comparably well with \emph{pathfinder}.

\begin{table}
\begin{tabular}{@{}llrrrr@{}}
\toprule
Annotator & Solver & \%Cov. & \%Used & Contigs & N50\\ \midrule
\multirow{2}{*}{GraphAligner}
& D-Wave & 80.18 (13.25) & 85.07 (13.49) & \textbf{1.10 }(0.54) &  9062 (1727)  \\
 & pathfinder & \textbf{93.24} (11.76) & 92.29 (11.00) & \textbf{1.10} (0.54) & \textbf{10060} (385)  \\ \midrule
\multirow{2}{*}{kmer2node} 
& D-Wave & 85.45 (11.31) & 91.52 \phantom{1}(8.55) & 1.97 (1.60) & 7203 (2315)  \\
& pathfinder & 85.99 (16.44) & 93.99  \phantom{1}(8.65) & 2.00 (1.59) & 7330 (2586) \\ \midrule
\multirow{2}{*}{minigraph} 
& D-Wave & 85.99 (10.85) & \textbf{94.75}  \phantom{1}(6.37) & 3.53 (1.71) & 4361 (1879) \\
 & pathfinder & 86.03 (11.15) & 94.43  \phantom{1}(6.93) & 3.53 (1.71) & 4492 (1913) \\ \bottomrule
\end{tabular}
\\[12pt]
\begin{tabular}{@{}llrrrr@{}}
\toprule
Annotator & Solver & Breaks & Indels & Diffs & \%Identity \\ \midrule
\multirow{2}{*}{GraphAligner}
& D-Wave &  5.77 (4.26) & 0.53 (0.96) & 0.07 (0.25) & 99.18 (0.98)    \\
 & pathfinder & 1.77 (2.73) & 0.83 (1.00) & 0.10 (0.30) & 99.22 (1.15)     \\ \midrule
\multirow{2}{*}{kmer2node} 
& D-Wave & 2.97 (2.34) & 0.50 (0.76) & 0.13 (0.34) & 99.08 (1.13)    \\
& pathfinder  & \textbf{1.23} (1.52) & 0.70 (0.90) & 0.13 (0.34) & 99.14 (1.03) \\ \midrule
\multirow{2}{*}{minigraph} 
& D-Wave & 1.30 (1.07) & \textbf{0.23 }(0.50) & \textbf{0.03} (0.18) & 99.55 (0.53)    \\
 & pathfinder & 1.27 (1.26) & 0.33 (0.70) & \textbf{0.03 }(0.18) & \textbf{99.58} (0.53)  \\ \bottomrule
\end{tabular}
\caption{Performance of pangenome guided mapping of short-read data using \emph{minigraph}, \emph{GraphAligner} and \emph{kmer2node} weight assignments with the \emph{D-Wave} and \emph{pathfinder} solvers. Averaged stats from 10 seeds with 3 alignments each, with standard deviations given in parentheses. \emph{D-Wave} was given 60 seconds per run, and 2 runs per alignment.}
\label{tab:dwave_solvers}
\end{table}

\subsection{Quantum optimisation}

We also test hybrid classical-quantum optimisation algorithms as a proof-of-concept.
We use various techniques to improve the convergence of the algorithm -- see~\cref{sec:meth_qaoa_sim} for details.
For the purposes of testing our algorithms, the majority of runs are performed with the quantum computer being simulated classically.
We can simulate circuits up to 35 qubits using heterogeneous CPU + GPU simulations.
However, we also perform small experiments on real quantum devices; a comparison of simulated and actual performance for one such experiment is provided below.

\begin{figure}
    \centering
    \includegraphics[alt={A pair of plots showing the convergence of the energy during a simulated QAOA run, and the samples drawn at the final iteration of the run.}]{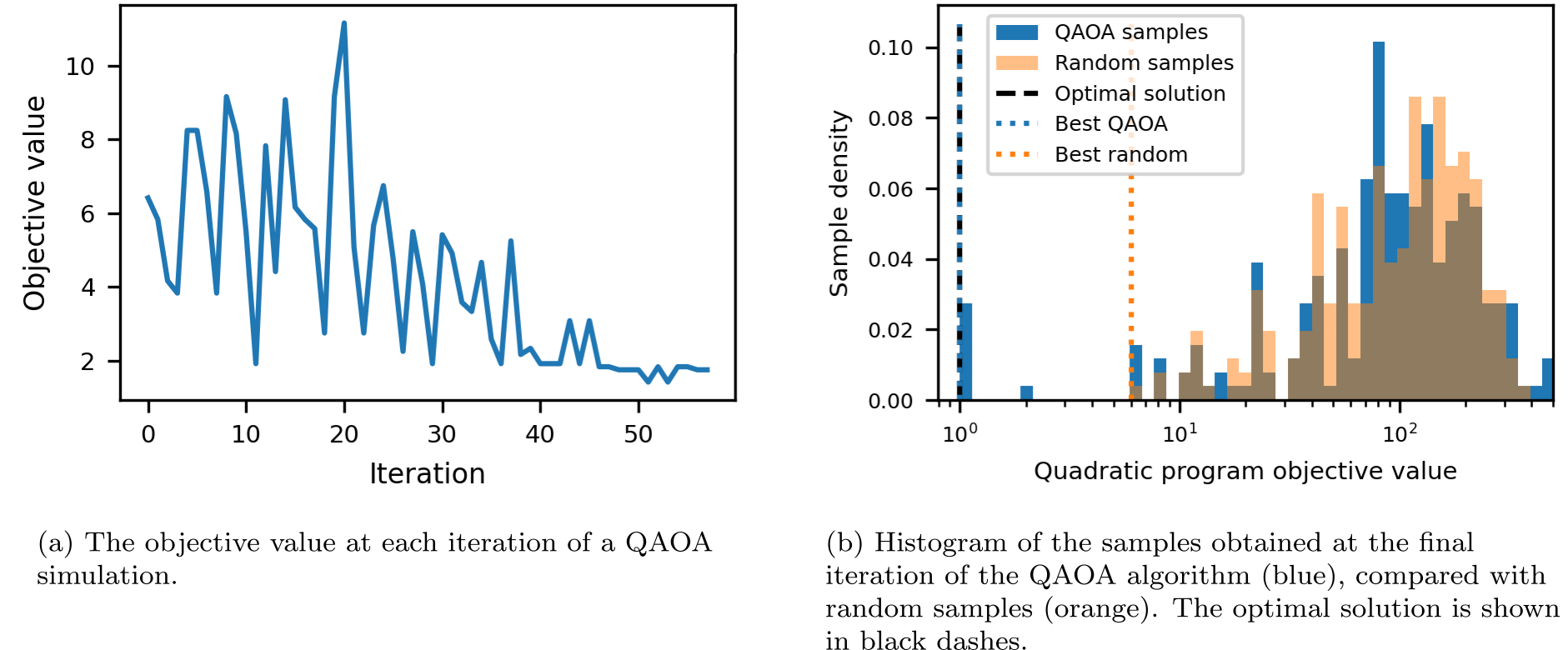}
    \caption{Plots showing the performance of a 10-qubit QAOA simulation for Oriented Tangle Resolution.}
    \label{fig:qaoa_performance}
\end{figure}

\Cref{fig:qaoa_performance} shows the results of simulating the QAOA algorithm with $p = 4$ on a small Oriented Tangle Resolution problem.
The circuit has 10 qubits, 496 2-qubit gates and a 2-qubit depth of 112 after being compiled to the IBM-native gate set.
The chosen emulator implements noiseless gates, but we include sampling noise: 256 samples are taken at each iteration.
A maximum of 100 iterations were allowed.
The simulation took under 20 seconds to run on a single CPU and GPU, including circuit compilation time.
\Cref{fig:qaoa_performance} demonstrates the successful convergence of the QAOA procedure.
The simulated quantum computer sampled the optimum repeatedly and had several other samples at the same energy as the best random samples.

\begin{figure}
    \centering
    \includegraphics[alt={A pair of plots showing the convergence of the energy during a hardware QAOA run, and the samples drawn at the final iteration of the run.}]{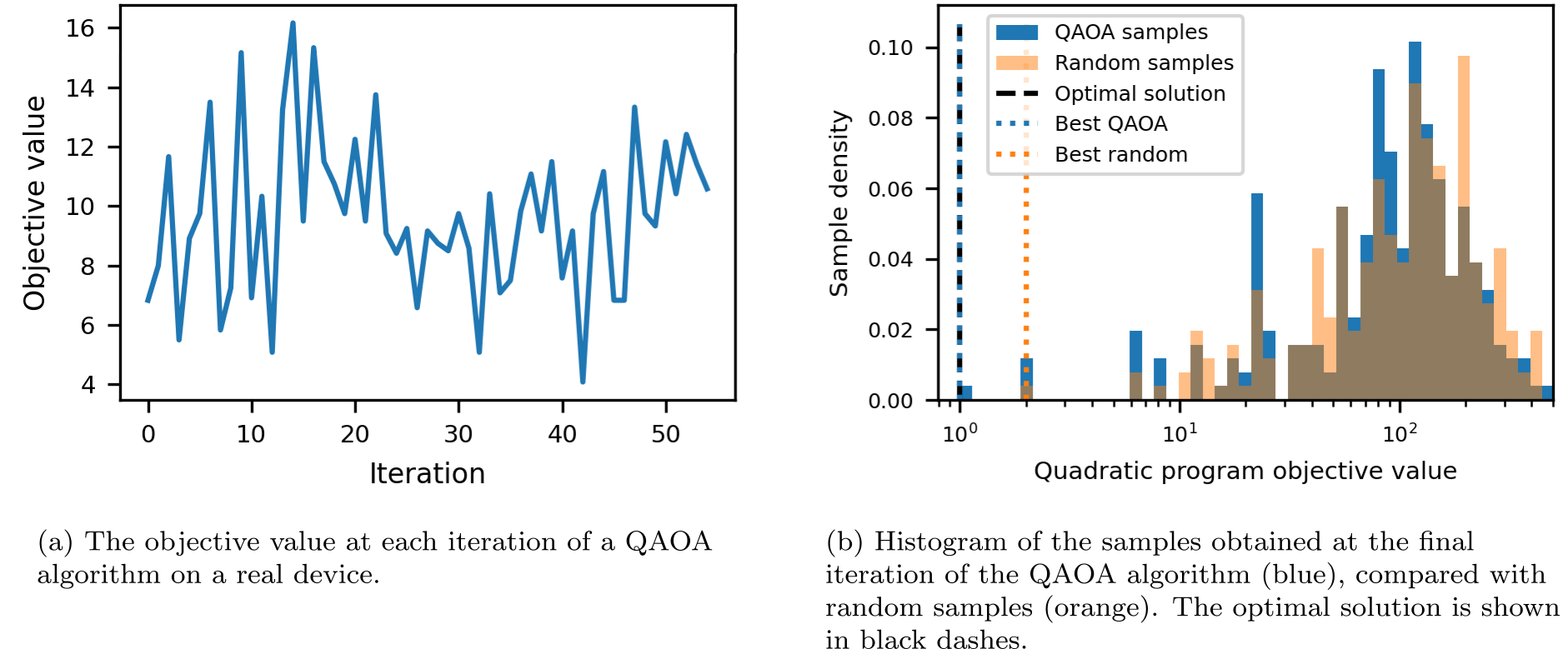}
    \caption{Plots showing the performance of a 10-qubit QAOA simulation for Oriented Tangle Resolution.}
    \label{fig:hardware_qaoa_performance}
\end{figure}

\Cref{fig:hardware_qaoa_performance} shows the results of running the same QAOA algorithm, with the same initial conditions, on an actual quantum device.
The device used is the \textit{IBM Strasbourg} computer, an Eagle R3 device with 127 qubits in a ``heavy-hex`` layout and an average 2-qubit error rate of around $3 \times 10^-2$.
The full computation took 213 seconds.
Each of the 55 iterations of the algorithm used less than 1s of QPU time; the remaining time is comprised of classical computation, communication with the device and preparing the device for the next iteration.
Clearly, noise in the device has degraded the convergence of the classical optimiser.
Nonetheless, \Cref{fig:hardware_qaoa_performance}b still shows a non-trivial shift of probability mass towards lower-energy solutions. 
Moreover, the optimal solution was again sampled, and several other low-energy solutions were found.

Going forward, it is clear that error mitigation techniques are critical for the QAOA algorithm to succeed.
Since we are essentially considering a sampling task, measurement error mitigation will likely be a good choice.
The implementation of these techniques is left for future work.

\section{Discussion}\label{sec:discussion}
In this work, we initiate a study on using pangenomes to aid in the assembly of short-read data.
We adopt a novel optimisation approach to tackle assembly, which we believe is well-suited to mitigate the noise inherent in short-read sampling.
We demonstrate that our methods enhance the accuracy of the assembly while significantly reducing the number of contigs reported compared to state-of-the-art \emph{de novo} assemblers.
The optimisation is designed to be amenable to both classical and quantum computing approaches and scales more favourably with problem size than existing methods.

Moreover, in challenging cases such as repetitive regions, our approach is more robust since errors during the copy number estimation step will be less impactful on our cost function approach compared to the existing state-of-the-art \textit{pathfinder}, whose copy number assignment is prescriptive.
For such instances, all sequence assembly methods will suffer from decreased accuracy due to the difficulty of assigning kmer counts; our \textit{kmer2node} tool includes mitigation strategies, which we discuss in~\cref{sec:kmer2node}.

Our pipeline consists of 3 main stages: graph annotation, copy number estimation and path finding.
We investigate the impact on solution quality of using different software for both the graph annotation and path finding steps.
For copy number estimation, we only use the capabilities of \textit{pathfinder}.
Further exploration in this direction could certainly be fruitful; one option might include training a machine-learning model to perform this task.

For the graph annotation, we find that \emph{minigraph} generally performs strongly, minimising the number of breaks in the assembly as well as the number of large indels, at the cost of reporting several contigs per assembly.
We usually prefer several accurate contigs over fewer, more inaccurate ones.

For the solvers, we see that our (classical) binary optimisation approaches are at the very least competitive with the bleeding-edge bioinformatics tool \emph{pathfinder}.
On the intermediate-size instances we test, the classical optimisers report good results within reasonably short time limits, often reporting better solutions than \emph{pathfinder} on specific problems.
They could be included as a parallel assembly method when high-quality results are desired, taking a consensus of all the candidate solutions as the final answer.
As the problem size increases and the exhaustive search of \emph{pathfinder} is no longer feasible, optimisation methods may be better placed to report reasonable assemblies.

We also investigate the feasibility of quantum computing in solving these problems.
We first test a QA approach on a set of middling-complexity problems, observing that the solution quality is broadly similar to \emph{pathfinder}, even though only short time limits were used.
This is a positive sign for the potential of using QA to solve industry-scale problems in the future, as hardware continues to improve.
However, there is currently no evidence of utility in using these systems for our problem.

Finally, we simulate the QAOA on some small problems as an initial indicator of the potential for standard, circuit-model quantum computers to solve pangenomic problems.
We can converge a QAOA experiment at this small size.
While performing a hardware experiment at industry scale is required to make any serious claims about quantum utility in this domain, we see this as a promising proof-of-principle. 

For a discussion of possible future research directions, see~\cref{sec:future_research}.

\section{Methods}
\subsection{Problem inputs}
Our primary input is unaligned short sequences of genomic data.
We have sufficient coverage of this to sequence every base in the target genomic region to a sufficient depth (typically at least 30x).
We also have a previously computed pangenome in Graphical Fragment Assembly (GFA) format, with DNA sequence fragments in graph nodes, and edges linking nodes together to describe paths through the graph corresponding to the training data.
The unaligned short sequences are then used to annotate this GFA with node and edge weights, where the weights indicate the expected number of traversals through those graph elements.
The problem is then how to identify the optimal path through the weighted graph that is most consistent with the specified weights.

In order to assess the accuracy of our methods, we use synthetic haploid data so we have a known ground truth.
We generate a population of 100 related genomes, iterated over 10 generations.
A random subset of 40 members of this population is used to build a pangenome, while a further 10 different genomes are used to evaluate our methods by simulating whole-genome shotgun experiments from a short-read sequencing instrument.
These short fragments are mapped back to the graph, producing the weighted GFA.

\subsection{Pangenome creation}


Tools such as \emph{minigraph} prefer to build a pangenome that is broadly co-linear with the input genomes.
This reduces the number of tangles and loops in the graph, but it means repeated sequence may end up in multiple nodes.
This is the style of pangenome published by the HPRC, which has the benefit of allowing standard linear reference coordinates, such as GRCh38, to be threaded through the pangenome graph.
Therefore, this is the style of pangenome graph we focus on.

A pangenome can also be complete or sparse: it could include every single base difference used in the input data, which leads to a large graph with many nodes and edges, or it could collapse small localised variations into a single node.
In this case, the node sequence will be either a representative single sequence or a consensus of all the donor genomes that passed through that node during graph construction.
\emph{Minigraph} takes this latter approach, which is useful since we need to keep the graph complexity low for the problem size to be amenable to existing quantum computers.

We build a simulated genome population using our own tool \emph{genome\_create} with fixed random seeds for reproducibility.
The initial genome contains Short-Tandem Repeats, longer CNVs, short and long repeat elements, translocations and inversions, and random point mutations.
Each subsequent population member is haploid, being derived from a single randomly chosen previous genome.
This is then further modified by the same types of variation listed above, but at a substantially reduced rate, giving rise to a population of similar sequences with a common complex structure that can form population graphs.
We use a subset of our population to build the pangenome graph using \emph{minigraph}.
The remaining sequences are used to simulate shotgun sequencing by replicating to 30-fold coverage and fragmenting into short sequences, with a uniform distribution of sequencing errors.

\subsection{Annotating the graph with copy numbers}

For each member of the training set, we produce fake short-read sequencing data.
For simplicity, this is single-ended (\ie not in pairs from a larger template).
They are randomly distributed from either strand with a random distribution of base-substitution errors; no indel errors are modelled.

Existing tools that take a short sequence, align it against a graph (GFA) and record the path include \emph{GraphAligner}, \emph{vg giraffe} and \emph{minigraph}.
We also produce our own \emph{kmer2node} tool.
All four tools annotate the GFA with edge-traversal counts and kmer counts within node sequences.
The presence of repeats can still be problematic, as multiple paths may exist.
It may not be possible to distinguish genuine sequence rearrangements from secondary hits arising from multiple copies of the repeated sequence in the pangenome.
When reported, we filter out secondary alignments.

As mentioned above, our pangenome graph node sequences are a single representative or consensus sequence.
However, the training genomes used to construct that pangenome may have many point mutations and small indels, which reduce kmer match rates. 
\emph{Kmer2node} may use a list of recognition sequences, so we align the full-length training genomes to the pangenome using \emph{GraphAligner} to produce a list of recognition sequences per node instead.
This improves our kmer hit rate, but may also decrease the percentage that uniquely map.
While indexing the GFA, we also get an estimate of the percentage of kmers we expect to be unique.
This provides us with a kmer hit-rate expectation during alignment of short-read data, allowing us to improve the assessment of node sequence depth.

All the tools produce a GFA with kmer counts per node and frequencies of edge traversals.
For \emph{pathfinder}, this is sufficient as it can convert these to depth-normalised counts.
To aid consistency, we also use \emph{pathfinder's} sub-graph connection analysis and depth-normalised counts as input to the QUBO solvers. 

\subsection{Classical post-processing: evaluating path solutions}\label{sec:classical_post}
Our path solvers produce a list of nodes and their orientation.
From this, we can concatenate the sequences recorded in the GFA nodes, reverse-complementing as necessary, to produce a candidate genome sequence.
Note that even if the path is perfect, this candidate genome may not be identical to the true genome, as our GFA files produced by \emph{minigraph} collapse small variations into a single node.
It is also possible that the true genome has novel structural rearrangements (copy-number changes or translocations) not present in the training data used to produce the pangenome graph, so a perfect path may not exist.

We align this candidate genome sequence against the known truth using \emph{bwa mem}, reporting only primary and supplementary hits.
This produces one or more alignments, each representing co-linear matches between the candidate and true genomes.
The alignments are broadly equivalent to the contigs produced by a \emph{de novo} assembler.
Each alignment record may, in turn, have smaller variations (substitutions and small indels) that are not large enough to split the alignment into two.
This provides a unified framework for comparing pangenome path reconstruction with more traditional \emph{de novo} sequence assembly tools.
We report the number of contigs and their N50 size.
We align the contigs to the true genome to estimate the proportion of the true genome we have sampled, and vice versa.
This two-way approach enables us to detect both over- and under-calling of copy-number variants.

These provide basic evidence on the contiguity of the assembled data, but not its accuracy.
To this end, we also report the number of breaks when aligning these contigs back to the true genome (indicating incorrect path forks or large indels that require a supplementary alignment).
These are analogous to false joins from a \textit{de novo} assembler.
For finer-grained evaluation, we also report small indels ($\geq$ 10 bp), regions of substantial base variation (multi-nucleotide polymorphism of 30\% variance within a 100 bp window), and the overall percentage identity.
Although BUSCO metrics are frequently used to assess assembly quality, these experiments assemble only small regions of synthetic genomes that do not contain any BUSCO genes, so it is not possible to use BUSCO for evaluation in this case.

We expect most users to prefer more contigs over incorrectly joined contigs, but the relative weighting of the two is a subjective decision that will depend on downstream usage.

A further optional refinement step is to realign the short reads to the generated candidate genome sequence and then take the consensus of the newly aligned sequences.
This corrects for novel mutations in this sample or simplifications made during the \emph{minigraph} pangenome creation step.

\subsection{Formulating sequence assembly as an optimisation problem}\label{sec:assembly2opt}

\begin{figure}
    \centering
    \includegraphics[alt={A schematic representation of the process of converting a graph-traversal problem to a QUBO problem, with sample QUBO constraint terms highlighted.}]{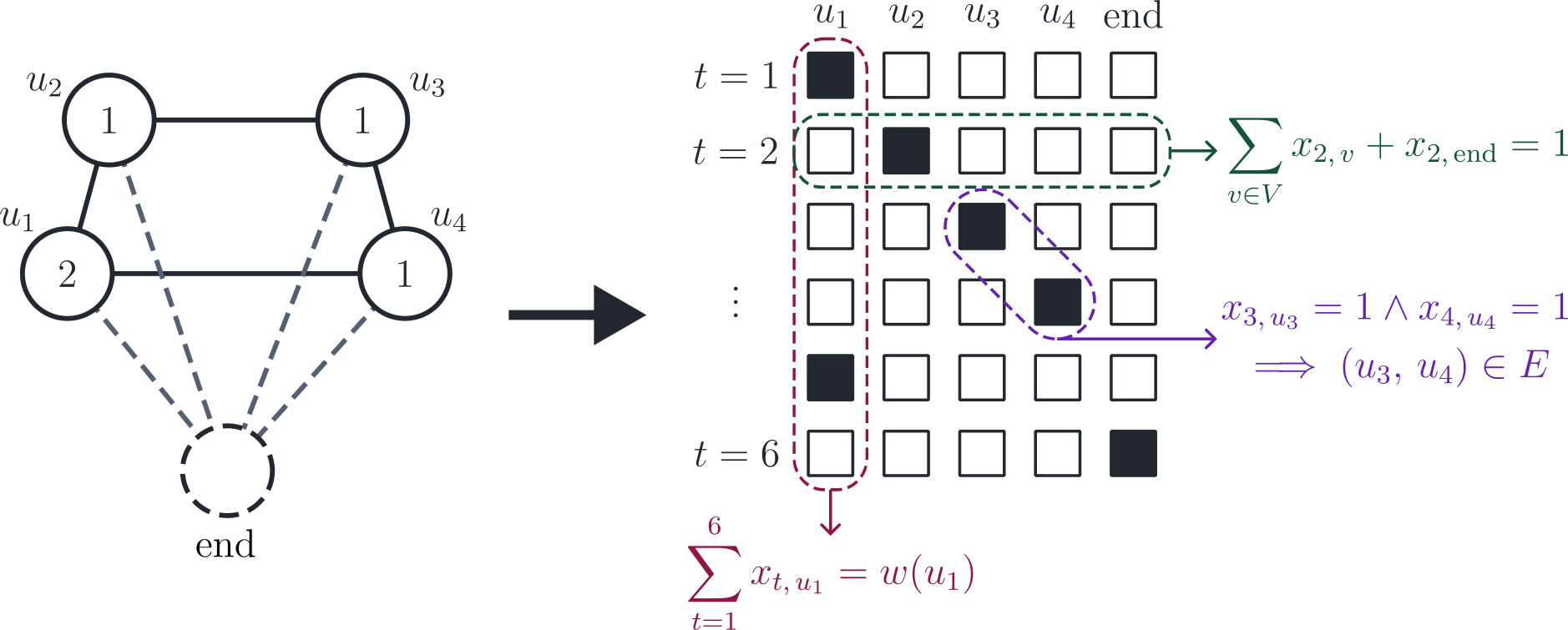}
    \caption{A sketch of the procedure for converting a graph-traversal problem to a QUBO problem. (Left) A graph annotated with weights, corresponding to a pangenome with reads aligned. Also shown is the virtual ``end'' node used in the QUBO formulation. (Right) A diagrammatic representation of the QUBO, with binary variables $x_{t,\,v}$ shown as squares that are filled if $x_{t,\,v}=1$. Highlighted are some of the constraints that we impose: in green, the constraint that a single variable is ``on'' at time 2; in purple, that the path traverses a graph step going from time 3 to 4; in red, that the path visits vertex $u_1$ twice, according to its weight.}
    \label{fig:graph_to_qubo}
\end{figure}

After annotation, our data is a graph with non-negative weights on the nodes and edges.
Our current formulation only uses node weights, whereas tools such as \emph{pathfinder} can use both sets of weights.
Extending our methodology to include edge weights is a future direction that we discuss in~\cref{sec:future_research}.

We formally define the pangenome graph annotated with copy numbers as a triplet
$G = (V, E, w)$,
where $V$ is the set of vertices, $E \subseteq V \times V$ is the set of edges and $w: V \rightarrow \mathbb{R}_+$ is a weight function giving the estimated copy number of each vertex.

We then seek a walk on the graph that best explains the copy number data.
This graph-traversal problem shares superficial similarities with the Travelling Salesman Problem.
In particular, we define the following optimisation problem, which we refer to as \emph{Tangle Resolution} due to the often complex, knot-like structures present in pangenomes.
Note the implicit requirement in the problem that each step in $W$ traverses an edge in $G$.\\
\begin{problem}[Tangle Resolution]\label{prob:tangle}
    Given a vertex-weighted graph $G = (V, E, w)$, find a walk $W$ on $G$ that minimises 
    \begin{equation}\label{eq:tangle}
        C_G(W) = \sum_{v \in V} \Bigl( \#W(v) - w(v) \Bigr)^2,
    \end{equation}
    where $\#W(v)$ is the number of times $W$ visits $v$.
\end{problem}

While this model serves as a useful starting point, it does not encompass all the biological features.
Most notably, it fails to account for the double-stranded nature of genomic data.
To better reflect the biology, we must consider a \emph{directed} graph.
The vertex set is now formed of 2 halves $V_+, V_-$.
The sequence data corresponding to $v_- \in V_-$ is the reverse complement of the sequence data corresponding to $v_+ \in V_+$.
Edges of the graph are directed and come in pairs: an edge $A+ \rightarrow B+$ implies the existence of an edge $B- \rightarrow A-$, and vice versa.
We call such a graph $G = (V_+\cup V_-, E, w)$ \emph{oriented}.

The corresponding optimisation task is similar, but we now sum the visits over both the positive and negative orientations.
This is captured in the following problem statement, which we refer to as \emph{Oriented Tangle Resolution}.\\
\begin{problem}[Oriented Tangle Resolution]\label{prob:oriented_tangle}
    Given a vertex-weighted, oriented graph $G = (V_+\cup V_-, E, w)$, find a walk $W$ on $G$ that minimises 
    \begin{equation}\label{eq:oriented_tangle}
        C_G(W) = \sum_{v \in V} \Bigl( \#W(v_+) + \#W(v_-) - w(v) \Bigr)^2,
    \end{equation}
    where $\#W(v_+), \#W(v_-)$ are the number of times $W$ visits $v_+$ and $v_-$ respectively.
\end{problem}

Finally, for the setting of human data, we need to resolve diploid DNA.
We therefore need to find a \emph{pair} of walks $W_1,\, W_2$ through an oriented graph that combine to explain the data.
Formally, we have \emph{Diploid Tangle Resolution}:\\
\begin{problem}[Diploid Tangle Resolution]\label{prob:diploid_tangle}
    Given a vertex-weighted, oriented graph $G = (V_+\cup V_-, E, w)$, find a pair of walks $W_1, W_2$ on $G$ that minimise 
    \begin{equation}\label{eq:diploid_tangle}
        C_G(W_1, W_2) = \sum_{v \in V} \left( 
        \sum_{i=1}^2  \Bigl( \#{W_i}(v_+) + \#{W_i}(v_-) \Bigr) - w(v) 
        \right)^2,
    \end{equation}
    where $\#W_i(v+), \#W_i(v_-)$ are the number of times walk $W_i$ visits $v_+$ and $v_-$ respectively.
\end{problem}
This formulation naturally extends to polyploid data by increasing the number of walks; we will not consider that setting here.

\subsection {Mapping guided alignment to binary optimisation}
The QUBO formulation defines a class of optimisation problems characterised by the minimisation of a quadratic polynomial over binary variables ${x_i}$. 
Since $x_i^2 = x_i$ for $x_i \in \{0,1\}$, any linear term can be replaced by a quadratic one.
Also, any constant offset can be neglected since it just shifts the energy landscape without affecting the optimal assignment.
Therefore, the objective can be formally defined as minimising $C(\{x_i\}) = x^TMx$, where $x\in \{0,1\}^n$ and $M\in \mathbb{R}^{n \times n}$ is a symmetric, real-valued square matrix.
For the sake of clarity, we express our cost functions here as standard polynomials, without reducing to a matrix form.
QUBO problems enjoy a high degree of expressivity, being able to represent classic combinatorial optimisation problems such as MAX-CUT and the Travelling Salesman Problem (TSP), while also remaining tractable to modern solvers at moderate sizes.
Solution techniques include generic branch-and-bound methods, as employed by commercial solvers such as \emph{Gurobi}, as well as domain-specific heuristics used by the open source library \emph{MQLib}.

QUBO is equivalent, under a linear change of variables, to the Ising model in statistical physics, a formulation extensively used in quantum annealing devices such as those developed by \emph{D-Wave} Systems. 
As such, QUBO is widely employed in quantum computing contexts, where its unconstrained nature and binary domain align with the hardware capabilities of current quantum and neuromorphic processors. 
The primary advantages of QUBO include its flexibility in encoding diverse problems within a unified mathematical framework and its compatibility with both quantum and classical heuristic solvers. Its unconstrained nature simplifies solver implementation by eliminating the need for explicit constraint handling, instead requiring constraint satisfaction to be incorporated into the objective function through penalty terms.

Notably, however, the QUBO forms of many classic graph-traversal problems, such as the Travelling Salesman Problem, require quadratically many variables in the graph size, with each variable encoding whether the path visits a certain node at a certain time.
QUBO formulations of these tasks traditionally use binary variables $x$ with 2 indices $t$ and $v$, with the variable $x_{t,v} = 1$ if the solution path visits vertex $v$ at time $t$.
The cost function is easily formulated by summing the cost of travelling between any 2 vertices $v$ and $v'$, $c_{vv'}$, over the edges that are traversed.
This leads to an objective term $\sum_{t, v, v'} c_{vv'} \cdot x_{t,v}x_{t+1,v'}$.
For an assignment to represent a valid path, it is necessary to impose the constraint that exactly one vertex is visited at each time.
This is accomplished through adding Lagrange multiplier terms to the cost function for each $t$, taking the form $\bigl(\sum_{v \in V} x_{t,v}- 1\bigr)^2$.

We propose a QUBO formulation for each of the 3 optimisation problems,~\cref{prob:tangle,,prob:oriented_tangle,,prob:diploid_tangle}.
For concreteness, we focus here on the simplest,~\cref{prob:tangle}.

Similarly to QUBO forms of the TSP, our formulation uses quadratically many variables.
In particular, our formulation uses $(N+1)T$ variables, where $N = |V|$ is the number of vertices in the graph and $T$ is the maximum length of a walk we are willing to consider.
In practice, we choose $T = \alpha \sum_{v \in V} w(v)$ for some constant $\alpha > 1$, which we usually take equal to $1.2$.
This allows walks to visit some vertices more than the data suggests, if doing so significantly improves the quality of another part of the solution.
The extra vertex introduced is a virtual ``end'' node with no sequence data attached.
A walk can spend several consecutive time steps in the end node without incurring any cost, allowing walks to effectively finish early.
However, any step that leaves the end node is heavily penalised.
A sketch of this process is given in~\cref{fig:graph_to_qubo}.

The total cost function consists of 3 parts: the first is a Lagrange multiplier term to enforce that the variable assignment encodes a walk that visits exactly 1 vertex at each time; the second is a multiplier term to enforce that, at each step, the walk traverses a graph edge, including travelling to or remaining in the virtual ``end'' node; the final term encodes the cost function $C_G(W)$.

In particular, we write the QUBO cost function as $C_G(\{x_{t,v}\}) = C^1_G(\{x_{t,v}\}) + C^2_G(\{x_{t,v}\}) + C^3_G(\{x_{t,v}\})$ where
\begin{align}
    C^1_G(\{x_{t,v}\}) &= \Lambda_1 
        \sum_{t=1}^T \left( \sum_{v \in V}x_{t,v} + x_{t,\text{end}}-1\right)^2,
    \\
    C^2_G(\{x_{t,v}\}) &= \Lambda_2 
        \sum_{t=1}^{T-1}\left( \sum_{v,v'in V}x_{t,v}x_{t+1,v'} \mathbbm{1}_{\{(v,v') \notin E \}} 
        + \sum_{v \in V}x_{t,\text{end}}x_{t+1,v}
        \right),
    \\
    C^3_G(\{x_{t,v}\}) &= 
    \sum_{v \in V} \left( \sum_{t=1}^T x_{t,v} - w(v) \right)^2.
\end{align}
There is some art to choosing the values of the Lagrange multipliers $\Lambda_1,\Lambda_2$.
If they are sufficiently large, then we can guarantee that the assignment that minimises $C_G(\{x_{t,v}\})$ encodes a walk that minimises $C_G(W)$.
However, if they are too large, then the energy landscape will have steep valleys around assignments that satisfy the constraints, and changes in the cost function will be nearly indistinguishable to the optimiser.
We find that intermediate-size values of $\Lambda_1 = 10$ and $\Lambda_2 = 5$ work well in practice.
Empirically, for instances where copy number estimation was accurate, the penalties are small enough that the $C^3_G$ part is significant enough for the optimiser to find the true solution.
Conversely, when copy number estimation is especially noisy, the penalties force a solution that obeys the constraints imposed by $C^1_G$ and $C^2_G$, overriding the inaccurate constraints of $C^3_G$.

To extend this method to Oriented Tangle Resolution, we simply introduce a pair of variables for each vertex and each time, resulting in $(2N+1)T$ variables.
For Diploid Tangle Resolution, we further introduce a new set of variables $y_{t,v}$ for the second path, resulting in $2(2N+1)T$ variables.
These variables have independent copies of $C^1_G$ and $C^2_G$ to enforce the constraints, but share a combined form of $C^3_G$.

\subsection{Mapping QUBO problems to quantum optimisation}
Quantum optimisation, especially for combinatorial optimisation problems, is a promising application of near-term quantum computers~\cite{kim_quantum_2025,mohanty_analysis_2023,perez-ramirez_variational_2024}.
One particularly fruitful technique is the Quantum Approximate Optimisation Algorithm (QAOA)~\cite{farhi_quantum_2014,blekos_review_2024,weidenfeller_scaling_2022,zhou_quantum_2020}.
In this setting, the cost function of the optimisation problem is cast to a quantum Hamiltonian, whose ground state encodes the optimal solution to the problem.

In the case of QUBO problems, this conversion is particularly simple.
The Hamiltonian $H$ has as many qubits as there are binary variables in the QUBO.
To obtain $H$, we replace each binary variable $x_i$ in the cost function $C$ with a quantum operator $(I-Z_i)/2$, where $Z_i$ is the Pauli-$Z$ gate acting on the $i^{\text{th}}$ qubit.
$H$ is then a sum over identity, $Z$ and $ZZ$ terms.
Since $Z\ket{0} = \ket{0}$ and $Z\ket{1} = -\ket{1}$, we see that $\tfrac{1}{2}(I-Z)\ket{b} = b\ket{b}$ for $b \in \{0,1\}$.
Therefore, the \emph{energy} of a state $\ket{x}$, given by $\bra{x}H\ket{x}$, corresponds exactly to the cost of the corresponding variable assignment $C(x)$.

QAOA is a hybrid classical-quantum optimisation process.
The outputs of a certain \emph{parameterised} quantum circuit are fed to a classical optimiser, which then suggests a new set of parameters.
This process is repeated until the classical optimiser terminates.
A final set of samples is taken from the circuit with the optimised parameters, and the sample with the best cost is taken as the solution to the optimisation problem.

The circuit is constructed from a layer of Hadamard gates followed by alternating \emph{cost} and \emph{mixer} layers.
The cost layer is $U_C(\gamma) = \exp(-i\gamma H)$.
There are many choices for a suitable mixer layer; we use $U_M(\beta) = \exp(-i\beta \sum_i X_i)$.
The number of such layers is a hyperparameter $p$.
In the limit as $p \rightarrow \infty$, QAOA approaches adiabatic quantum computing, which is known to be universal for quantum computing~\cite{aharonov_adiabatic_2005}.
In contrast, for $p=1$, QAOA is known to be classically simulable under certain circumstances~\cite{hastings_classical_2019}.
In the absence of noise, increasing $p$ should increase the convergence rate of QAOA.
In practice, the effects of noise mean that a moderate $p$ value is most likely to be effective.

\subsection{Classical simulation of QAOA}\label{sec:meth_qaoa_sim}
We perform a thorough classical simulation of the QAOA algorithm for Tangle Resolution.
We primarily use the \emph{Qiskit} library, including the GPU-enabled \emph{Aer} simulation package.
Simulations are run using up to 4 Nvidia H-100 GPU units and up to 512 GiB of RAM.
The \emph{Qiskit} library enables heterogeneous GPU/CPU simulations at large scales, enabling full statevector simulations up to 35 qubits on our hardware. 

All of our simulations contain shot noise, \ie at each iteration, a finite number of samples are taken from the probability distribution according to the final statevector.
The classical optimiser is then given only these samples from which to estimate the progress of the optimisation.

We find that the convergence of our experiments is greatly improved by using Conditional Value at Risk (CVaR) as the accumulation function, instead of a more traditional energy minimisation.
In this setting, only the top-performing samples are taken into consideration when evaluating the merit of the QAOA parameters.
CVaR only cares about ``high value'' samples with low energy, and high energy outliers are discarded.
Recent studies have justified CVaR both analytically~\cite{barkoutsos_improving_2020} and empirically~\cite{barkoutsos_improving_2020,kolotouros_evolving_2022,cai_enhancing_2025}.


\begin{framed}
    \textbf{Key points}
    \begin{itemize}
        \item Using pangenomes to resolve short-read data avoids linear reference bias that is unavoidable in traditional reference-guided mapping.
        \item Compared to \emph{de novo} assembly, pangenome-guided assemblies report far fewer contigs and cover a greater amount of the true genome, with only a small increase in the number of false joins.
        \item Our novel optimisation-based approach achieves comparable results to bleeding-edge exhaustive methods, and will remain feasible as problem size increases.
        \item Our methods are designed to run efficiently on quantum computers as quantum technology improves; already, quantum annealing methods achieve good results on moderate-sized instances.
    \end{itemize}
\end{framed}

\section*{Acknowledgements}
The authors would like to thank the entire QPG team for their valuable support and discussions throughout the process of this work. 

\section*{Funding}
Work on the Quantum Pangenomics project was supported by Wellcome Leap as part of the Q4Bio Program. 
JC was supported by the EPSRC.
JB was additionally funded by the Wellcome Trust [Grant number 220540/Z/20/A].
SS was also supported by the Royal Society University Research Fellowship.

\section*{Data availability}
All code used to produce data and plots is available on GitHub at \hyperlink{https://github.com/jkbonfield/qpg}{https://github.com/jkbonfield/qpg}.

\section*{Biographical note}
Josh Cudby is a PhD student at the University of Cambridge and a Research Assistant at the University of Oxford. His research focuses on quantum algorithms for bioinformatics.


\printbibliography

\pagebreak
\SupplementaryMaterials
\begin{center}
\textbf{\Large Supplementary Information}
\end{center}
\section{A primer on quantum computing and quantum optimisation}\label{sec:quantum}

Quantum computing (QC) is an emergent technology that uses the principles of quantum mechanics to perform certain tasks faster than is possible on traditional, classical computers.
For example, quantum computers can find the prime factors of large numbers exponentially faster than the best classical algorithms known to computer science.
Key physical properties that QC exploits are \emph{superposition} and \emph{entanglement}.
Superposition refers to the ability of quantum bits, or \emph{qubits}, are able to exist in both the ``0'' and ``1'' state \emph{at the same time}. 
Intuitively (and extremely informally!), this opens the door for quantum speed-ups by allowing functions to be computed on many inputs simultaneously.
Entanglement is a phenomenon in which the state of each particle in some collection cannot be described independently of the others.
Acting on one particle in the collection then leads to the ``spooky action at a distance'' described by Einstein.

A quantum process is most commonly described in the quantum circuit model.
In analogy with classical computing circuits, a certain input state is acted on by a sequence of quantum logic gates, which are usually restricted to act on at most 2 qubits.
At the end, readout is achieved by taking a measurement of some or all of the qubits.
Quantum measurements are probabilistic, outputting a ``0'' or ``1'' with probabilities in accordance with the state before the measurement was taken.
The output of a quantum circuit is therefore simply a bit string, identical to those produced by classical circuits.
The end goal of quantum algorithms is to move as much probability mass as possible onto bit strings that encode the correct answer for a given input.

While the potential for quantum advantages is appealing, there are downsides to the use of QC, at least currently.
Quantum devices are inherently noisy and error-prone.
The downside of exploiting superposition is that quantum states are analogue signals, rather than the digital state of classical bits.
Interference from the environment on the physical system that contains the quantum computer causes the quantum states to drift.
Algorithms often rely on constructive or destructive interference to shift probability onto desired states, and this drift reduces how effectively that process works.
Also, quantum logic gates are prone to errors, with 2-qubit gates failing at a rate of around $4 \times  10^{-4}$ on leading hardware.
While certain error mitigation strategies are possible, and quantum error correction is promising in the long run, errors are unavoidable on current hardware.
Moreover, quantum computers are limited in size, with no more than 156 qubits present on commercially available hardware.
The presence of noise and the lack of qubits mean that quantum algorithms for current hardware must be designed to be resource-efficient.
No more than a few thousand gates can be run without the combination of errors and noise destroying any information.

There are several promising avenues for algorithms that might experimentally demonstrate the utility of quantum computers on current or near-term hardware.
Arguably chief among them is quantum optimisation and the poster-child algorithm, the Quantum Approximate Optimisation Algorithm (QAOA).
QAOA is a hybrid classical-quantum algorithm, consisting of a quantum circuit whose gates are parametrised by a set of variables, and a classical optimiser.
The quantum circuit is sampled several times, and these samples are used to estimate a cost function.
The cost function is then used by the classical optimiser to suggest new parameters for the quantum circuit.
The process is iterated until the optimiser reports convergence or some other termination criterion, such as the total number of samples, is reached.
QAOA experiments have been successfully run with up to 127 qubits.

Although quantum circuit model hardware is currently limited, there is an intermediate technology that allows for preliminary experimentation.
Quantum Annealing (QA) is an optimisation process that can directly be applied to combinatorial optimisation problems.
Quantum annealers exist outside of the quantum circuit model and are not able to perform general computational tasks.
However, they are well-suited to the problems we consider in this work.
Moreover, implementations of QA, most notably by \emph{D-Wave}, are much more advanced than their circuit counterparts.
\emph{D-Wave}'s latest system has over 4,400 qubits, allowing classically-challenging problems to be tackled by quantum hardware.
While there is some debate as to the validity of these systems as true quantum computers, we view them as a useful indication of the future power of QC.

\section{Future research directions}
\label{sec:future_research}
It is tempting to draw comparisons and make projections about how indicative \emph{D-Wave}’s success might be for the potential advantages of QAOA methods. However, there is no strong evidence to suggest that successful performance on a \emph{D-Wave} machine provides any meaningful insight into the performance of QAOA on the same QUBO formulation. A thorough study of this connection remains out of reach due to the limited number of qubits and the shallow gate depths available on current circuit-based quantum platforms. Moreover, for this problem, running QAOA with a QUBO Hamiltonian may not be the most efficient approach, as the required number of qubits scales unfavourably with the size of the problem instance. Instead, one should consider methods that bypass the QUBO formulation altogether and encode the objective directly into the Hamiltonian in a fundamentally different way.

One drawback of the current optimisation formulation is the lack of edge information.
When we have graphs with multiple paths through the same set of nodes, it can be impossible to distinguish nodes A+ B+ C+ from A+ B- C+.
\emph{Pathfinder} combines edge weights with node weights to resolve these conflicts.
Experiments using~\emph{pathfinder} with and without edge weights have shown that using this information has a minor impact on most statistics, but can reduce the number of false joins by up to a third.

While it is straightforward to include an edge weight term in the abstract optimisation cost functions of~\cref{eq:tangle,eq:oriented_tangle,eq:diploid_tangle}, it is not clear how to translate such terms into a QUBO.
The information of whether an edge was traversed at a certain time is inherently quadratic, given by terms $x_{t,\,i}x_{t+1,\,j}$.
To obtain a cost function that is minimised when the number of traversals of an edge is close to the weight of that edge requires terms like $\sum_t \big(x_{t,\,i}x_{t+1,\,j} - w(i,\,j)\big)^2$ for each $(i,\,j) \in E$.
These higher-order terms incur overheads in the number of variables if they are re-expressed as quadratic terms.
A local graph preprocessing step to remove zero-weight edges that do not break graph connectivity is likely to achieve an approximate equivalent without the need to incorporate edge data directly into the QUBO model.

All graph path finding methods are impacted by the quality of the input kmer counts.
Further improvements can be made to these tools, particularly concerning non-unique mapping, which would improve the quality of all solvers.
However, there is further missing information yet to be considered, where larger sequence fragments trace a path through many small nodes, forming longer-range constraints rather than the nearest neighbour ones encoded in the graph edge counts.

The kmer mapping software used to generate node weights does not use Illumina read-pairing metrics.
In a traditional linear reference alignment, the read pairing is an important factor for resolving some repeats, and the same could also be applied to pangenome mapping.
Pairing information could potentially also be used in inferring paths, utilising knowledge of the template size distribution.

A major drawback of the QUBO formulation is the $\mathcal{O}(NT) = \mathcal{O}(N^2)$ scaling in the number of variables, a nearly quadratic increase compared to the natural $\mathcal{O}(T\log(N)) = \mathcal{O}(N\log(N))$ number of variables required to encode a length-$T$ walk on $N$ nodes in binary.
While such overheads are permissible in a complexity-theoretic sense, practically, they can lead to huge increases in the difficulty of finding optimal solutions.
Moreover, the number of qubits required essentially precludes industry-scale problems from being tackled with our current approach on a quantum computer in the near term.

One possible direction that improves resource usage and also allows edge weight data to be included is the use of Higher-order Unconstrained Binary Optimisation (HUBO) variants. 
HUBO formulations are more expressive than QUBO, with terms containing products of arbitrarily many binary variables; the downside is that (classically) solving them is generally more difficult and less well-explored in the literature.
Notably, however, HUBO may not have such a solution overhead when mapped to QAOA algorithms on quantum computers.
In that setting, each layer of the cost Hamiltonian now contains multi-qubit $Z$ rotations rather than single- and two-qubit rotations only.
These gates need to be compiled down to 2-qubit gates, but constructions that are linear in the number of qubits involved exist~\cite{glos_space-efficient_2020}.
In this setting, the inclusion of edge weight data requires the implementation of more multi-qubit interactions, which in turn increases the circuit depth.
Notably, there are no overheads in the number of qubits, which is a major limiting factor in current-term quantum hardware.
We could also consider Prog-QAOA~\cite{bako_prog-qaoa_2025} formulations, which directly encode optimisation problems into QAOA cost Hamiltonians, without going through a QUBO or HUBO step.
We expect that either of these 2 encodings could achieve $\mathcal{O}(N\log(N))$ qubit scaling, at the cost of increased circuit depth.

Another potential improvement is to modify the cost function to place greater emphasis on larger nodes.
Larger nodes have less variance in the number of unique kmer hits, so we should have more confidence that the annotated values represent the true copy number.
Also, they represent a larger part of the underlying sequence, so they are more important to get correct.
For example, we might modify the objective of Tangle Resolution, given in~\cref{eq:tangle}, to
    \begin{equation}\label{eq:tangle_new}
        C_G(W) = \sum_{v \in V} \log(L(v)) \Bigl( \#W(v) - w(v) \Bigr)^2,
    \end{equation}
where $L(v)$ is the length of the sequence stored in node $v$.

We will also investigate the performance of the QAOA solvers using hardware experiments rather than simulations.
On hardware with up to 156 qubits and an encoding that uses only $\mathcal{O}(N\log(N))$, we could encode problems with around 30 nodes.
These are large enough to be classically non-trivial for certain hard instances, and would provide strong evidence for quantum utility in this domain if the quantum algorithms performed well.

Finally, we will properly test and benchmark the performance of the Diploid Tangle Resolution solvers.
In this setting, it is more subtle to evaluate the quality of the solution, as the two paths might erroneously swap which allele they are representing.
We will look to incorporate methodology from existing phasing software to aid in solution refinement and evaluation.

\section{Details of node weight assignment methods}\label{sec:node_weight}

A critical component of the robustness of path finding is the quality of the input node coverage scores.
These are processed by the solvers to yield a node copy-number, but the GFAv1 specification permits a kmer count in the \texttt{SC:i} tag.
Depth can be converted to coverage by dividing by the length (or expected observable length), and depth to coverage by observation of average depth across nodes or knowledge of the input sequencing depth.

\subsection{Minigraph}\label{sec:minigraph}

The \emph{minigraph}~\cite{li_design_2020} option \texttt{--cov} may be used to redisplay the input GFA file with a report of the percentage of kmers observed in the \texttt{dc:f} tag.
This is modified to the standard kmer-count \texttt{KC:i} tag by multiplying by the node length.
For edges, \texttt{dc:f} is converted to the \texttt{EC:i} tag.

This also permits visualisation within the \emph{Bandage} tool and for normalisation with other tool chains.

The command line options for synthetic short read data are:

\begin{verbatim}
minigraph -x sr -j 0.01 --cov $gfa reads.fa > reads.mg
perl -we script-to-add-tags reads.mg > reads.gfa
\end{verbatim}

\subsection{Kmer2node}\label{sec:kmer2node}

The basic strategy of \emph{kmer2node} is to build a kmer index of the sequence in each node and then to match every kmer in each short read against this computed index.

\begin{figure}[t]
Input genomes to build the graph, with spacing for visualisation purposes only:
\begin{verbatim}

    CAT ACTCC CCC GGACAAGG TAG
    CAT ACACC CCC GGTCATGG TAG
    CAT GGTGG CCC  ACATT   TAG
    CAT GGAGG CCC  ACTTT   TAG

\end{verbatim}
\includegraphics[width=0.95\linewidth,alt={A schematic representation of the kmer2node annotation strategy on a small graph.}]{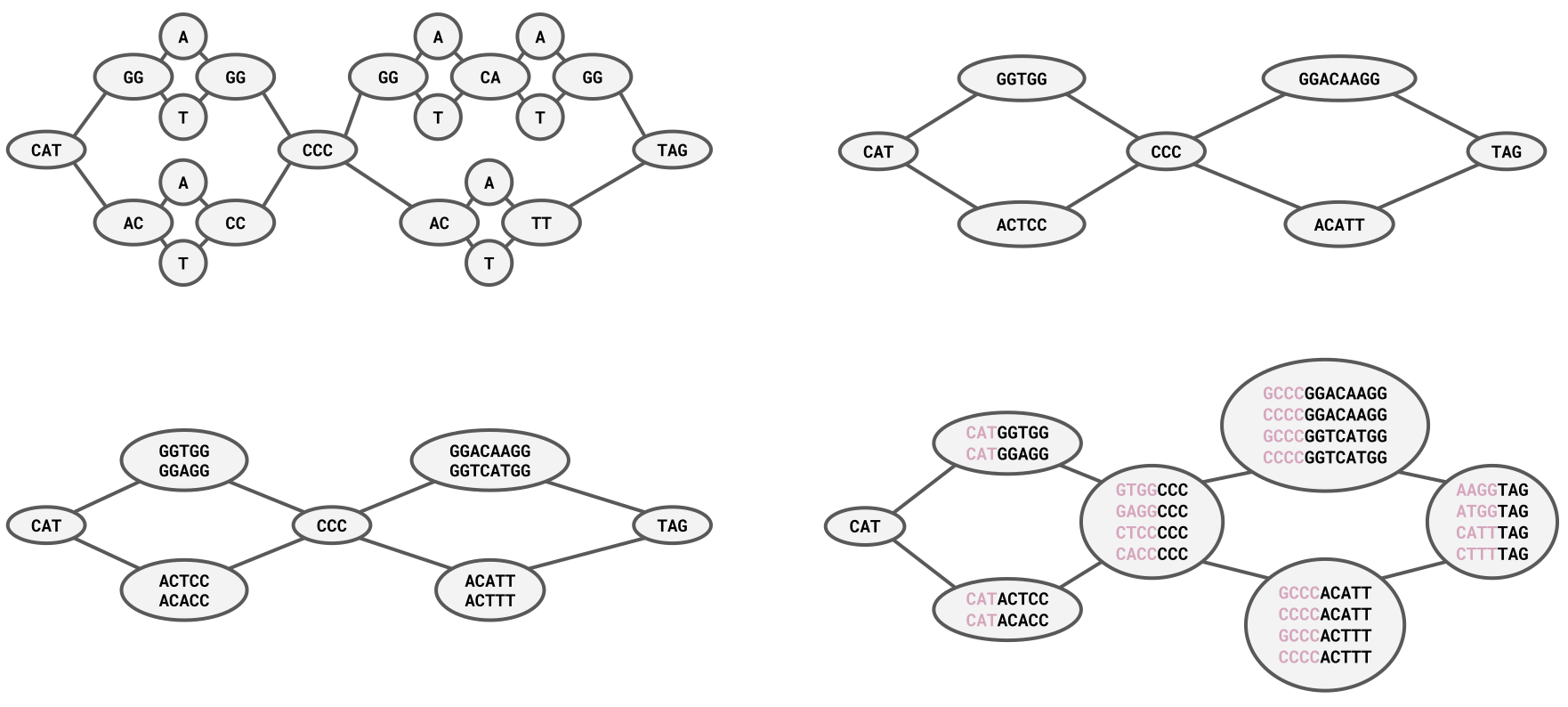}
\caption{A pictorial view of \emph{kmer2node} processing on a graph built from 4 short example genomes.  On the top left we have a fully complete pangenome capturing variation at the individual base level (\emph{minigraph-cactus} style).  Top right is a lossy graph with small variants collapsed into representative nodes (\emph{minigraph} style).  Bottom left is produced by aligning the full length genomes used to build the graph back to the graph and tracking which sequence substrings align to each node.  Bottom right is after recursively walking back through the graph from each node to generate $k-1$ prefix bases. }
\label{fig:kmer2node}
\end{figure}

The pangenome graphs we process come from \emph{minigraph}, which does not produce new nodes for minor sequence variation, leading to imprecise mapping.
This is beneficial as it reduces the complexity of the graphs to resolve, making the problem more amenable to the quantum environment, but it also means kmer matching may miss some hits.
We do, however, have the original input genomes that were used to build the pangenome.
By aligning these back to the pangenome graph with \emph{GraphAligner}, we can get a path of node names along with a CIGAR string.
Combined, these yield the exact sequence from the training genomes that covers each node rather than the single one reported by \emph{minigraph}.
This ancillary ``nodeseq'' file is also indexed, improving the mapping accuracy of \emph{kmer2node}.
See~\cref{fig:kmer2node} for a pictorial demonstration of these steps.

The repetitive nature of the sequences in the graph nodes means that not all kmers can be uniquely placed.
A unique node sequence of length $L$ will have $L - k + 1$ kmers of length $k$.
However, during indexing, we track which kmers are unique and which are not to record an expected number of unique kmers per node sequence.
This is used during mapping to compensate for non-unique segments.
Additionally, a sequence of kmers from a short-read that spans unique to non-unique and back to unique within the same graph node means we can still identify the internal non-unique kmers as belonging to that node.
We can additionally rescue some non-unique kmers at the start and ends of nodes by a similar strategy.
Finally, short reads that transition from one node to another node with non-unique kmers may still be disambiguated if the graph edges indicate only one combination of nodes is possible.

At present, \emph{kmer2node} does not exploit knowledge of paired Illumina reads, but, in principle, these can also be used to resolve some repeat locations.
Improved handling of repetitive kmers should also permit mapping qualities to be produced, providing partially qualified evidence for more paths.
For now, \emph{kmer2node} restricts itself to being conservative in reporting matches.

For this work, the command lines used are:

\begin{verbatim}
GraphAligner -g $gfa -f train.fa -x vg -a pop.gaf
for k in 75 35 20;do
    gaf2nodeseq2.pl pop.gaf train.fa $gfa $k > $gfa.nodeseq.$k
    kmer2node4 -G $gfa -U -k$k $gfa.nodeseq.$k reads.fa -E /dev/stdout > reads.$k
done
merge_kmer2node.pl $gfa reads.k* | tag_gfa_km.pl $gfa > reads.gfa
\end{verbatim}

The tool was written purely for research into this problem and is not yet considered production quality.
It also maps every kmer and is considerably slower than comparable minimiser techniques.
It can be found in the qpg tangle GitHub repository along with the evaluation code.

\subsection{GraphAligner}\label{sec:graphaligner}

\emph{GraphAligner}~\cite{rautiainen_graphaligner_2020} cannot emit kmer count data.
However, the standard alignment format does include an alignment path indicating the nodes used, along with a CIGAR string indicating which strings of bases map against which node.
Nodes internal to a multi-node path string (e.g. ``s2'' and ``s3'' in ``\textgreater s1 \textgreater s2 \textgreater s3 \textgreater s4'') will have their kmer count incremented by the length of the node.  Nodes on the end of the path string have their kmer count incremented according to the relative proportion of the remaining sequence.
This is used to fill out \texttt{KC} and \texttt{SC} auxiliary tags on GFA ``S'' nodes and \texttt{EC} on ``L'' edges.

Note \emph{GraphAligner} cannot distinguish between primary and secondary alignments.
A sequence may align to multiple locations without regard to which is best.
The \emph{tag\_gfa\_ga.pl} program attempts to identify the best single alignment, but this may still be a non-uniquely mapped repeat region.
This gives rise to higher average node coverage than \emph{minigraph} and \emph{kmer2node}, but with poorer specificity.
This appears to help \emph{pathfinder}, which does not like many zero-depth nodes, but hampers the QUBO formulations.

The parameters used for running GraphAligner are:

\begin{verbatim}
GraphAligner --min-alignment-score 90 --seeds-minimizer-ignore-frequent 1e-2 \
    -f $gfa -f reads.fa -x vg -a reads.gaf
tag_gfa_ga.pl $gfa 10 reads.gaf > reads.gfa
\end{verbatim}

\subsection{Giraffe}\label{sec:giraffe}

Unlike the other tools used vg's \emph{giraffe}~\cite{siren_pangenomics_2021} aligner is a dedicated short-read aligner, capable of understanding Illumina read pairs.
However, our synthetic data is not currently simulated read-pairs.

\emph{Giraffe} does not emit kmer data, but it can be used to produce the same GAF format as \emph{GraphAligner}.
We then use the same pipeline to process the GAF node strings into fake kmer data.
One key difference, however, is \emph{giraffe} creates robust mapping qualities, making it easier to select the best aligned copy when multiple alignments are reported, and to filter out poor matches.

The command lines for running \emph{vg giraffe} are:

\begin{verbatim}
vg autoindex --workflow giraffe --prefix $gfa -g $gfa
vg giraffe -Z $gfa.giraffe.gbz --named-coordinates -f read.fa -o gaf > read.gaf
tag_gfa_ga.pl $gfa 10 reads.gaf > reads.gfa
\end{verbatim}

\section{Preliminary results of graph edge preprocessing}\label{sec:trim-edges}

The annotation methods can provide coverage data for both nodes and edges, with edge data being produced when we observe a single sequence of kmers transitioning from one node to another.
A significant source of sequence ``breaks'' arises from inversions where a sequence has been observed in the pangenome to be flipped in orientation. meaning we have a graph where a node can be traversed in either orientation.
Only edge information can be used to decide which orientation should be utilised.
However, the QUBO formulations do not currently take into account edge weights, which puts them at a disadvantage compared to \emph{pathfinder}.
While a formal inclusion of edge weights is possible, it adds complexity to the model.

We implemented a local graph preprocessing step that observes cases where a node has multiple input and output edges and at least one input and one output has non-zero edge weight and non-zero node weight.
In this case, any additional edges with zero weights can be deemed to be superfluous and removing them will not break graph connectivity.  The results can be seen in ~\cref{tab:edge-trim}, which demonstrates a significant drop in the number of breaks and a slight increase in percentage covered and used.

\begin{table}[ht]
{\begin{tabular}{@{}llrrrrrrr@{}}
\toprule
Annotator & Trim & \%Covered & \%Used & No.Contigs & No.Breaks & No.Indel & No.Diff & \%Identity \\ \midrule
kmer2node    & N & 86.5 & 91.6 & 2.6 & 2.4 & 0.5 & 0.3 & 98.7 \\
kmer2node    & Y & 87.0 & 92.6 & 2.6 & 2.1 & 0.6 & 0.3 & 98.7 \\ \midrule
minigraph    & N & 84.2 & 91.7 & 4.4 & 2.5 & 0.4 & 0.2 & 99.2 \\
minigraph    & Y & 84.1 & 92.1 & 4.3 & 2.0 & 0.4 & 0.1 & 99.2 \\ \midrule
GraphAligner & N & 87.2 & 77.2 & 1.6 & 5.9 & 0.9 & 0.3 & 98.4 \\
GraphAligner & Y & 89.4 & 78.0 & 1.6 & 5.4 & 0.9 & 0.3 & 98.5 \\ \midrule
giraffe      & N & 87.6 & 76.2 & 1.6 & 6.7 & 1.0 & 0.2 & 98.6 \\
giraffe      & Y & 88.5 & 78.5 & 1.6 & 5.8 & 1.0 & 0.3 & 98.5 \\ \bottomrule
\end{tabular}}
\caption{Results of the MQLib solver running for 30 seconds on 51 graphs aligning 5 sequences each, with and without edge trimming enabled.  Results are with the consensus optimisation applied.}
\label{tab:edge-trim}
\end{table}

\section{Preliminary results of hybrid assembly and kmer mapping}\label{sec:hybrid-kmer}

Some synthetic problem examples demonstrate \emph{de novo} assembly as the most accurate genome reconstruction, while others suggest that k-mer mapping techniques are the best.
We can run \emph{syncasm} to produce an assembly graph with sequence in the nodes.
Each node is a conservatively assembled contiguous sequence (a contig), with the edges holding possible alternatives caused by either multiple haplotypes or uncertainty arising from repeated sequences.
Using the assembled contigs as input to the pangenome mapping (normalised for our expected depth), optionally alongside the raw sequence fragments, can potentially give us the best of both prior solutions.

Preliminary evidence in ~\cref{tab:hybrid-asm-map} demonstrates a modest improvement in accuracy, particularly when combined with \emph{kmer2node}.

\begin{table}[ht]
{\begin{tabular}{@{}llrrrrrrr@{}}
\toprule
Annotator & Input & \%Covered & \%Used & No.Contigs & No.Breaks & No.Indel & No.Diff & \%Identity \\ \midrule
kmer2node    & R    &   88.4  &  94.0  &  2.6   &  1.5  &  0.6  &  0.3  &  98.7 \\
kmer2node    & C    &   85.3  &  93.7  &  3.3   &  1.6  &  0.4  &  0.2  &  99.0 \\
kmer2node    & R+C  &   89.4  &  94.0  &  2.3   &  1.5  &  0.6  &  0.3  &  98.6 \\ \midrule
GraphAligner & R    &   93.1  &  92.7  &  1.6   &  1.6  &  0.8  &  0.3  &  98.6 \\
GraphAligner & C    &   90.8  &  93.8  &  2.2   &  1.5  &  0.6  &  0.2  &  98.9 \\ 
GraphAligner & R+C  &   93.2  &  92.6  &  1.6   &  1.7  &  0.8  &  0.3  &  98.6 \\ \midrule
syncasm      & C    &   91.6  &  86.2  & 31.3   &  0.0  &  0.1  &  0.0  & 100.0 \\ \bottomrule
\end{tabular}}
\caption{Demonstration of the impact of combining \emph{de novo} assembly and kmer mapping techniques.  The input column lists ``R'' for raw reads and ``C'' for contigs produced from running \emph{syncasm} on the raw reads.  For \emph{kmer2node} we show that raw data and contig data alone are poorer than the combination of the two, while with \emph{GraphAligner} there is minimal change. We also include an assessment of \emph{syncasm} directly without using the pangenome.  \emph{Kmer2node} figures are from the optimised consensus.}
\label{tab:hybrid-asm-map}
\end{table}

\end{document}